\documentstyle[aps,psfig]{revtex}
\newcommand{\be}{\begin{equation}}
\newcommand{\ee}{\end{equation}}
\newcommand{\half}{\frac{1}{2}}
\newcommand{\bea}{\begin{eqnarray}}
\newcommand{\eea}{\end{eqnarray}}

\newcommand{\x}{\vec{x}}

\newcommand{\bphi}{\bar{\phi}}
\newcommand{\bpi}{\bar{\pi}}
\begin{document}
\preprint{ITFA-97-32}
\draft
\title{Real-time propagator for high temperature dimensional reduction}
\author{B. J. Nauta and Ch. G. van Weert}
\address{Institute for Theoretical Physics, University of Amsterdam\\
        Valckenierstraat 65, 1018 XE Amsterdam, The Netherlands}
\date{19-8-1997}
\maketitle
\begin{abstract}
We discuss the extension of dimensional reduction in thermal
field theory at high temperature to real-time correlation
functions. It is shown that the perturbative corrections to the
leading classical behavior of a scalar bosonic field theory are 
determined by an effective
contour propagator. On the real-time-branch of the time-path
contour the effective propagator is obtained by subtracting
the classical propagator from the contour 
propagator of thermal field theory, whereas on the Euclidean 
branch 
it reduces to the non-static Matsubara propagator of
standard dimensional reduction. 
\end{abstract}
\pacs{PACS number: 11.10.Wx}

\section{Introduction}

It has been argued that the dominant non-perturbative behavior of bosonic 
quantum fields at high temperature may be described by a classical effective 
theory \cite{grigoriev}.
The basic idea is to replace the full quantum theory by a scheme in which 
one solves classical equations of motion with initial conditions taken from a 
thermal ensemble.

In this letter we discuss the derivation of this classical approximation from 
the time-path formulation of thermal field theory. The thermal state is
taken into account by extending the time evolution along a contour in the
complex \(t\)-plane as shown in fig. 1. The contour consists of three parts: 
a forward time branch $C_1$ on the real axis, a backward branch $C_2$ and
a third branch $C_3$ down the Euclidean path \(t=t_{in}-i\tau\) from 
\(\tau=0\) to 
\(\tau=\beta\), with \(\beta=T^{-1}\) the inverse temperature \cite{lebellac}. 
The classical limit is obtained in the stationary phase approximation.
On the real-time contour \(C_{12}=C_{1}\cup C_{2}\) this yields the 
classical equations of 
motion, whereas on the the Euclidean branch the fields must be taken 
stationary and equal to the initial values for the real time evolution 
\cite{bodeker}.

A systematic improvement involves the calculation of the effective action, 
both on the Euclidean, and the real-time contours. 
This is a generalization of the approach of dimensional reduction on the Euclidean contour,
in which one constructs an effective 3D-theory for the static 
Matsubara mode. The bare parameters of the 3D-theory are matched to the 
corresponding physical parameters in the full thermal field theory
\cite{aarts1,aarts2}, by a
perturbative calculation of the static two- and four-point functions \cite{kajantie,jakovac}.

Below we present a scheme in which this procedure is generalized to 
the real-time contour. We show that the non-static Matsubara propagator
generalizes to a contour 
propagator which is the difference of the contour propagator of thermal 
field theory and the classical propagator. The non-classical 
effective action on 
the real-time contour has loop contributions determined by this generalized 
non-static contour propagator.

\section{Classical limit}
In the time-path formulation of thermal field theory \cite{lebellac},
real-time correlation functions may be obtained from the generating functional
\be
Z[j]=\int D\phi D\pi\; \exp{iS[\phi,\pi]+j\cdot\phi}\;.
\label{genfun}
\ee
The dot-notation is an abbreviation for
$j\cdot\phi=i\int_{C}dt d^3x j(x)\phi(x)$.
We consider the simple model of a scalar theory in Minkowski space 
\(x=(t,\vec{x})\), with action
\be
S[\phi,\pi]=\int_{C}dt\int d^3x\left[\pi\dot{\phi}-\half\pi^2-
\half(\nabla\phi)^2-\half m^2\phi^2-\frac{1}{4!}\lambda\phi^4
\right]\;,
\ee
parameterized by a bare mass \(m\) and a coupling constant \(\lambda\),
and we assume that the system is prepared
in thermal equilibrium at some initial time \(t_{in}\). 
The time-contour \(C\) is depicted in fig 1.
Occasionally we shall indicate fields on different parts of the contour as 
\(\phi_{r}(t,\vec{x})=\phi(t,\vec{x})\), \(t\in C_{r}\; r=1,2,3\).
In order that 
the KMS condition is satisfied, the fields are periodic with respect to the 
begin- and end-point of the contour:
\(\phi_{1}(t_{in},\vec{x})=\phi_{3}(t_{in}-i\beta,\vec{x})\).

In dimensional reduction one seeks to obtain from the generating functional
(\ref{genfun}) an effective theory for the static mode
\({\cal P}\phi(x)=\Phi(\vec{x})\) and its conjugate momentum
\({\cal P}\pi(x)=\Pi(\vec{x})\), where the projection operator is defined as
\be
{\cal P}\phi(x)=iT\int_{C_3}dt\phi(t,\vec{x})
=T\int_{0}^{\beta}d\tau\phi(t_{in}-i\tau,\vec{x})\;.
\label{projector}
\ee
The effective theory is obtained by separating off the integration over the static fields and writing
\be
Z[j]=\int D\Phi D\Pi \exp{W[\Phi,\Pi;j]}\;.
\label{genfun2}
\ee
The effective action of the dimensionally reduced theory is formally given by
the constrained generating functional
\be
W[\Phi,\Pi;j]=\log \int D\phi D\pi\, \delta\left({\cal P}\phi -\Phi\right)
\delta\left({\cal P}\pi-\Pi \right)\exp{iS[\phi,\pi]
+j\cdot\phi} \; ,
\label{effact}
\ee
which contains the contributions of the non-static modes. 
The difference with the constraint effective potential defined in
\cite{raifeartaigh}, is that the fields 
\(\Phi(\vec{x}),\Pi(\vec{x})\) still depend on the spatial coordinates. 
The effective action in (\ref{effact}) is a direct generalization
of the dimensionally reduced action for a scalar field; 
see e.g. \cite{jakovac}. Indeed, 
if the time contour is confined to the Euclidean branch
the fields may be decomposed into Matsubara modes
\(
\phi(t_{in}-i\tau,\vec{x})=T\sum e^{-i\omega_{n}\tau}
\phi(i\omega_{n},\vec{x})\;,\)
with \(\omega_{n}=2\pi n T\) the Matsubara frequencies. This simplifies 
the integration measure in (\ref{effact}) to one over the non-static modes
\(n\not=0\).

We proceed with (\ref{effact}), and perform a shift 
\(\phi(x)\rightarrow \bar{\phi}(x)+\phi(x)\);
\(\pi(x)\rightarrow \bar{\pi}(x)+\pi(x)\). The background fields
\(\bar{\phi}(x)\) and \(\bar{\pi}(x)\) are continuous extensions of the 
fields \(\Phi(\vec{x})\) and \(\Pi(\vec{x})\), that is, they are 
equal to  \(\Phi(\vec{x})\) and \(\Pi(\vec{x})\) on \(C_{3}\), but arbitrary 
on the contour \(C_{12}=C_1 \cup C_2\), except for the initial conditions 
\(\bar{\phi}(t_{in},\vec{x})=\Phi(\vec{x})\), \(\bar{\pi}(t_{in},\vec{x})=
\Pi(\vec{x})\) required by periodicity.

It is convenient to separate off all terms in the shifted action
\(S[\bar{\phi}+\phi,\bar{\pi}+\pi]\) that are of second-and-higher order in 
the fields \(\phi(x)\) and \(\pi(x)\). This defines a new action
\be
\bar{S}[\phi,\pi]=S[\bar{\phi}+\phi,\bar{\pi}+\pi]
+i\delta_{\bar{\phi}}S[\bphi,\bpi]\cdot\phi
+i\delta_{\bar{\pi}}S[\bphi,\bpi]\cdot\pi-S[\bphi,\bpi]\;.
\label{sbar}
\ee
The term linear in $\pi(x)$ can be made to vanish by imposing
\be
\delta_{\bar{\pi}}S[\bar{\phi},\bar{\pi}]=\dot{\bphi}-\bpi=0\;,
\quad t \in C_{12}\;.
\ee
On the Euclidean branch the contribution of this term vanishes on 
account of the fact that the 
background fields are time independent there and the constraints 
on the integration in (\ref{effact}). For the same reason the term
linear in $\phi$ on the right-hand side of (\ref{sbar})
gives no contribution on $C_3$. 
As a result of these manipulations we find for the constraint action 
defined in (\ref{effact})
\be
W[\Phi,\Pi;j]=iS[\bar{\phi},\bar{\pi}]+\bar{W}[J]+j\cdot\bar{\phi}\;.
\label{effact2}\ee
The second term is entirely determined by the shifted action (\ref{sbar})
\be
\bar{W}[J]=\log \int D\phi D\pi \delta\left(\cal{P}\phi \right)
\delta\left(\cal{P}\pi \right) \exp{i\bar{S}[\phi,\pi]+J\cdot\phi}\;.
\label{wbar}
\ee 
The source has the value
\(
J=j+\delta_{\bar{\phi}}S[\bar{\phi}] 
\) 
on the real part \(C_{12}\) of the contour, where 
\(S[\bar{\phi}]=S[\bar{\phi},\dot{\bar{\phi}}]\) is 
the ordinary action of a \(\lambda\phi^4\)-theory. 

If we identify the background fields with the solution to the 
stationary phase approximation for (\ref{genfun}), the source \(J\) 
is equal to zero,
 and we are left with the first and last term on the right-hand side of 
(\ref{effact2}). 
This constitutes the classical approximation of thermal field theory
\cite{bodeker,aarts1,aarts2,buchmuller}.
The time evolution of the field \(\bar{\phi}(x)\) is then determined by 
the classical equations of motion \(j+\delta_{\bar{\phi}}S[\bar{\phi}]=0\) 
in Minkowski space. 
The remaining functional integral in (\ref{genfun2}) averages over the initial conditions with thermal weight \(\exp iS[\Phi,\Pi]=\exp -\beta H[\Phi,\Pi]\).

\section{quantum corrections}
The interaction of the classical background field \(\bar{\phi}(x)\) with
the quantum and thermal fluctuations is contained in the generating 
functional (\ref{wbar}). These quantum corrections will be determined 
perturbatively. 
We insert an integral representation \(
\delta\left({\cal P}\pi\right)=\int D\chi\exp\chi\cdot\pi
\), for the momentum delta-function in
(\ref{wbar}).
The auxiliary field \(\chi(x)\) is zero everywhere except on \(C_{3}\)
where it has a spatial dependence:
\(\chi(t_{in}-i\tau,\vec{x})=\chi(\vec{x})\).
Interchanging the order of integration, we 
first integrate out the momentum variable \(\pi(x)\) and
subsequently the auxiliary field
\(\chi(\vec{x})\). The relevant Gaussian integral is
\be
\int D\chi \int D\pi \exp{i\int_{C}dt d^3 x
\left[-\half \pi^2+(\chi+\dot{\phi})\pi\right]}
=\exp {\left(\half \dot{\phi} \cdot \dot{\phi}
-\half {\cal P}\dot{\phi}\cdot {\cal P}\dot{\phi}\right)}\;,
\label{intoutpi}
\ee
where the second factor comes from the constraint on the momentum integration.
Recalling definition (\ref{projector}) for the projector, we deduce
\be
{\cal P}\dot{\phi}(x)=iT\int_{C_3}dt \dot{\phi}(x)
=iT[\phi_{1}(t_{in},\vec{x})-\phi_{2}(t_{in},\vec{x})]\;.
\label{difference}
\ee
The boundary condition imposes periodicity on the fields over the whole 
contour \(C\), 
but not on the Euclidean segment \(C_{3}\) separately. Hence, the integral 
(\ref{difference}) gives a finite contribution
which is equal to the difference of the field values at the begin and 
end points of the real-time contour \(C_{12}\); see fig 1.

Substituting the result (\ref{intoutpi}) into (\ref{wbar}), we obtain 
\be
e^{\bar{W}[J]}=\int D\phi \delta \left(\cal{P}\phi \right)
\exp{\left( i\bar{S}[\phi] - \half {\cal P}\dot{\phi}\cdot {\cal P}\dot{\phi}
+J\cdot\phi\right)}\; .
\label{wbar2}
\ee
In this expression the action $\bar{S}[\phi]
=S_0[\phi]+\bar{S}_I[\phi]$ is the ordinary action for a scalar 
\(\lambda\phi^4\)-theory translated by a background field \(\bar{\phi}(x)\), 
with quadratic part \(S_{0}[\phi]\)
and interaction part 
\be
\bar{S}_{I}[\phi]=\frac{1}{4}
\begin{array}{c}\mbox{\psfig{figure=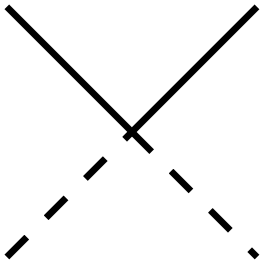,height=0.5in,angle=-90}}\end{array}
+\frac{1}{3!}
\begin{array}{c}\mbox{\psfig{figure=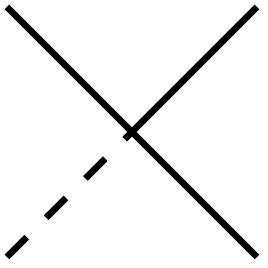,height=0.5in,angle=-90}}\end{array}
+\frac{1}{4!}
\begin{array}{c}\mbox{\psfig{figure=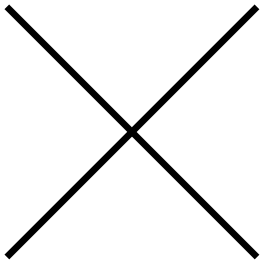,height=0.5in,angle=-90}}\end{array}\;.
\label{intact}\ee
The striped lines denote the background field \(\bar{\phi}\)
and the solid lines denote the quantum field \(\phi\).
The interactions are treated in the usual perturbative manner 
\be
\bar{W}[J]=\left.
e^{i\bar{S}[-i\delta_{J}]}e^{\bar{W}_{0}[J]}\right|_{con}\;.
\label{wbar=con}\ee
So we can focus our attention on the free part of the generating functional
(\ref{wbar2}). Using again an auxiliary field, we write
\be
e^{\bar{W}_{0}[J]}=e^{-\half {\cal P}\dot{\phi}\cdot{\cal P}\dot{\phi}}
\int D\chi\int D\phi e^{iS_{0}[\phi]+(J+\chi)\cdot\phi}\;.
\label{wbarnul}\ee
We have brought out the term quadratic in
\be
{\cal P}\dot{\phi}(x)=T\left[ \frac{\delta}{\delta J_{1}(t_{in},\vec{x})}-
\frac{\delta}{\delta J_{2}(t_{in},\vec{x})}\right]\;.
\label{correction2}\ee
The Gaussian 
integral over the field \(\phi\) can be performed and leads to
\be
e^{\bar{W}_0[J]}= Z_0[0] e^{- \half {\cal P}\dot{\phi}\cdot {\cal P}\dot{\phi}}
 \int D\chi\exp\half (J+\chi)\cdot D_C \cdot (J+\chi)\; ,
\label{wnul}
\ee
with \(D_{C}(x-x')\) the well-known free contour propagator of thermal field theory 
\cite{lebellac}. In the next section we will show by explicitly evaluating 
(\ref{wnul})
that the free generating functional is a standard bilinear functional of \(J\)
as for free fields
\be
e^{\bar{W}_{0}[J]}={\cal N}e^{\half J\cdot \Delta_{C}\cdot J}\;,
\ee
but with a new contour propagator
\be
\Delta_{C}(x-x')=D_{C}(x-x')-S_{C}(x-x')\;,
\label{newprop}\ee
where \(S_{C}(x-x')\) turns out to be the propagator of the classical theory 
on the real-time contour and the static Matsubara propagator on \(C_3\).

\section{contour propagator}
In this section we calculate the free action \(\bar{W}_{0}[J]\) 
given in (\ref{wnul}). 
Writing out the Gaussian exponent, 
we encounter the 3D propagator \(\tilde{D}_{3D}=1/\omega_{k}^{2}\)
(the static Matsubara propagator), 
and the time-integrated propagator
\be
\tilde{f}_{C}(t,\vec{k})= iT\int_{C_{3}}dt'\tilde{D}_{C}(t-t',\vec{k})=
\left\{
\begin{array}{ll}
   \frac{T}{\omega_{k}^2}\cos\omega_k(t_{in}-t)& \; t \in C_{12}\\
   \frac{T}{\omega_{k}^2} &\; t\in C_3
\end{array} \right. ,
\ee
with \(\omega^{2}_{k}=\vec{k}^2+m^2\). Here we used the explicit 
expression for the contour propagator in \cite{lebellac}, formula (3.74).
By the usual procedure of shifting the integration variable, we obtain
the result 
\be
e^{\bar{W}_0[J]}= Z_0[0] e^{- \half {\cal P}\dot{\phi}\cdot {\cal P}\dot{\phi}}
e^{\half J\cdot D_C^{{\rm I}} \cdot J} \; ,
\label{wnul1}
\ee
with a subtracted contour propagator
\(
D_{C}^{{\rm I}}(x,x')=D_{C}(x-x')- S_C^{{\rm I}}(x,x')\),
\be
S_C^{{\rm I}}(x,x')=\beta\int \frac{d^3 k}{(2\pi)^3} e^{i\vec{k}\cdot(\x-\x')}
\omega_{k}^2\tilde{f}_C(t,\vec{k})\tilde{f}_C(t',\vec{k})\; .
\label{corprop1}
\ee
Note that this expression is not time-translation invariant.

We will now show that the exponential factor quadratic in
\({\cal P}\dot{\phi}(\x)\) in eq.(\ref{wnul1})
leads to a further subtraction of the contour propagator.
Using (\ref{correction2}) we first calculate
\be
\half {\cal P}\dot{\phi}\left(\,J\cdot D^{{\rm I}}_{C}\cdot J\right)= 
-T\int d^4x'\left[
D_{1s}^{{\rm I}}(t_{in},t')-D_{2s}^{{\rm I}}(t_{in},t')\right]J_{s}(x') \; ,
\ee
with \(D_{rs}(t,t')=D_{C}(t,t')\) with \(t \in C_{r}, t'\in C_{s}\).
We suppressed the spatial dependence of the propagators. Their difference
is easily calculated. On the Euclidean contour it vanishes.
Using the Fourier transform we find for \(t'\in C_{12}\)
\be
\tilde{g}_{C}(t',\vec{k})=
\tilde{D}^{{\rm I}}_{1s}(t_{in}-t')-\tilde{D}^{{\rm I}}_{2s}(t_{in}-t')
= \frac{i}{\omega_{k}}\sin\omega_{k}(t_{in}-t'),\;\;\; s=1,2 .
\label{bstep1}
\ee
The subtraction (\ref{corprop1}) gives no contribution. Next we notice the 
property
\be
\half {\cal P}\dot{\phi} {\cal P}\dot{\phi}\,
\left(J\cdot D_{C}^{{\rm I}}\cdot J\right)=0 \; .
\label{bstep2}
\ee
The two results (\ref{bstep1}) and (\ref{bstep2}) together imply that the 
first exponent 
\(-\half {\cal P}\dot{\phi}\cdot{\cal P}\dot{\phi}\) 
in (\ref{wnul1}) may be replaced by
\(-\half J\cdot S^{{\rm II}}_{C}\cdot J\), 
with the propagator
\be
S_{C}^{{\rm II}}(x,x')=T \int \frac{d^3 k}{(2\pi)^3}e^{i\vec{k}\cdot(\x-\x')}
\tilde{g}_{C}(t,\vec{k})\tilde{g}_{C}(t',\vec{k})\;.
\label{corprop2}
\ee
The integrand is given by (\ref{bstep1}) for \(t\in C_{12}\), and is zero for 
\(t\in C_{3}\). 

This all leads to the following conclusion: 
the subtraction term \(S_{C}(x-x')\) in (\ref{newprop}) is given 
by the sum of (\ref{corprop1}) 
and (\ref{corprop2}). 
Writing this out on the various parts of the contour, we find
\be
\tilde{S}_C(t-t',\vec{k})=\left\{
\begin{array}{ll}
\frac{T}{\omega_{k}^2}\cos\omega_k(t-t')& t,t'\in C_{12}\\
\frac{T}{\omega_{k}^2}\cos\omega_k(t_{in}-t')& t\in C_{3}, t'\in C_{12}\\
\frac{T}{\omega_{k}^2}\cos\omega_k(t- t_{in})&t\in C_{12}, t'\in C_{3}\\
\frac{T}{\omega_{k}^2}&t,t'\in C_{3}
\end{array}
\right.
\label{clasprop}
\ee
We recognize the first term in this list as the familiar classical 
free two-point 
function \cite{parisi}. The last term is the static Matsubara propagator.
The second and third term connect the vertical Euclidean time branch \(C_{3}\) 
with the real-time branch \(C_{12}\). In
thermal field theory it is shown that in the limit $t_{in}
\rightarrow -\infty $, the two real-time contours
decouple from the Euclidean branch, provided that the
infinitesimal damping coefficient $\epsilon$ in the spectral
density is kept finite till the end of the
calculations \cite{evans,landsman}.  

\section{Conclusions and outlook}

We have argued that for the case of a scalar bosonic field, the scheme
of dimensional reduction may be extended to real-time correlation functions.
Like in imaginary-time dimensional reduction, the IR behavior of the 
effective theory is improved because on internal lines
classical modes are excluded. 
This allows a perturbative calculation of the quantum
corrections to the effective action
\be
\Gamma[\bar{\phi}]=iS[\bar{\phi},\bar{\pi}]+\Gamma_{\bar{\phi}}[0]\;.
\ee
The Feynman rules for calculating the loop contributions in the background
field \(\bar{\phi}(x)\) are the usual ones of thermal field theory, save for 
the fact that solid lines connecting the vertices in eq(\ref{intact}), 
represent 
the contour propagator (\ref{newprop}). If the external background field is 
conveniently chosen to be equal to the expectation value of the quantum field,
the background field satisfies the equation of motion
\be
\frac{\partial \Gamma[\bar{\phi}]}{\partial \bar{\phi}(x)}=j(x)\;,
\label{effeqmot}\ee 
on the real-time contour. The equation of motion (\ref{effeqmot}) has to 
be solved with the initial conditions 
\(\bar{\phi}(t_{in},\vec{x})=\Phi(\vec{x})\), 
\(\dot{\bar{\phi}}(t_{in},\vec{x})=\Pi(\vec{x})\).
This may be compared with the effective action as obtained 
by Greiner and Muller
\cite{greiner} by integrating out hard modes, \(\vec{k}^2>k^2_{c}\sim 
\lambda T^2\). As a consequence their vertex functions are cut-off dependent.

As an example let us consider the case of linear relaxation for a given
external source \(j(x)\) which is the same on both branches. This leads to 
the equation of motion \cite{lawrie}
\be
-\partial^2\bar{\phi}(x)+m^2\bar{\phi}(x)+\int d^4x'\Sigma(x-x')\bar{\phi}(x')=
j(x)\;,
\ee
where \(\Sigma(x-x')\) is the real-time retarded self-energy on \(C_{1}\)
and the advanced self-energy on \(C_2\), both calculated with the contour 
propagator (\ref{newprop}). Compared to the explicit expression for the 
retarded self energy to one-loop-order in the full quantum theory 
\cite{aarts2,lawrie}, the
zero-mode contribution is seen to drop out. This ``matches'' the mass of the 
quantum theory to the mass of the effective classical theory
\(
m^2\rightarrow m^2+\lambda T^2/24-(c.t.)_{3D}\;.
\)
The last term is the counterterm for the 3D-theory that cancels the 
linear divergence in the one-loop two-point function.
A calculation of the coupling constant for the effective theory gives 
the standard dimensional reduction
 result \cite{jakovac}.
So we find an effective theory which, in the local approximation and to 
one-loop-order, reduces to the classical theory with the matching relations 
as proposed in \cite{aarts1,aarts2,buchmuller}.

\newpage
\begin{center} {\large FIGURE}
\vspace{1cm}
\end{center} 
\centerline{\psfig{figure=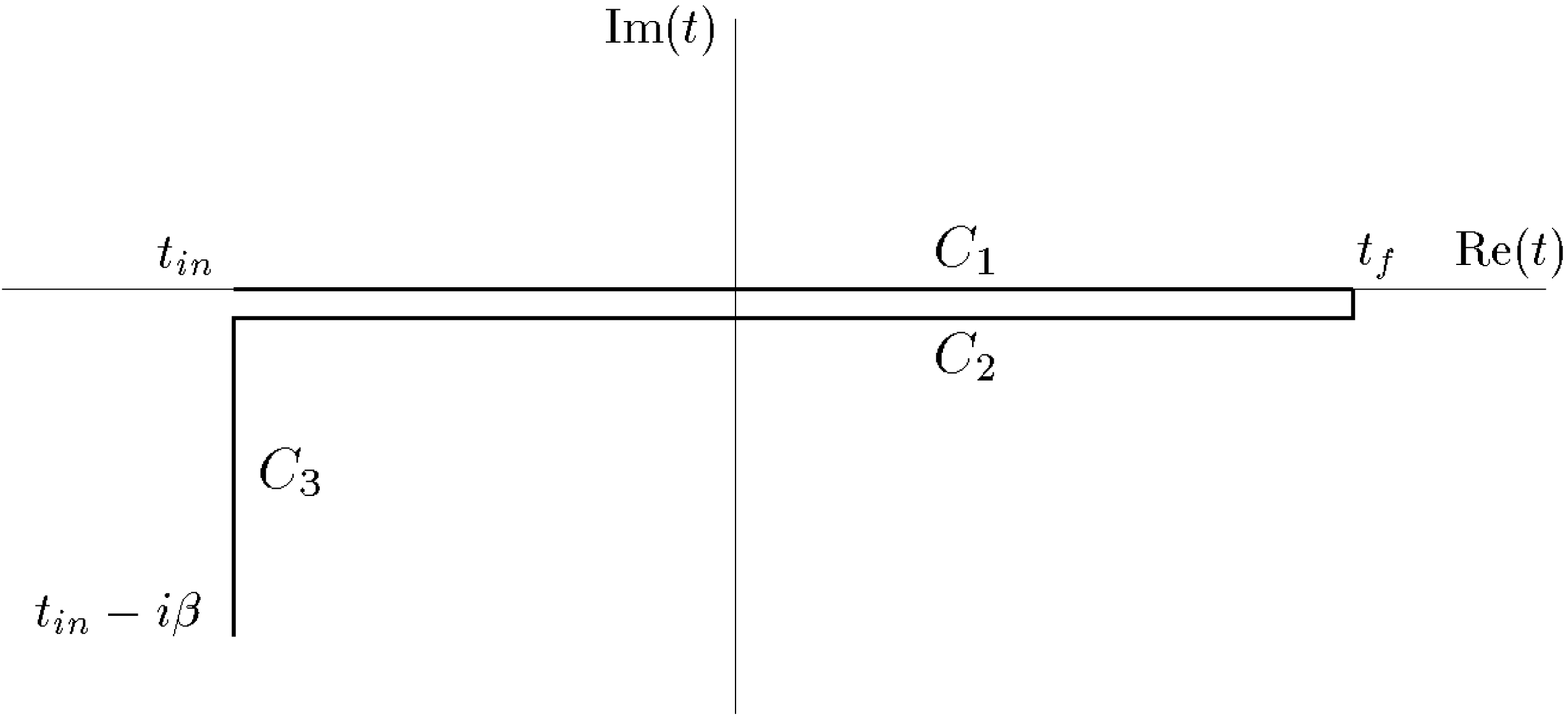,height=2.5in,angle=-90}}
\vspace{0.5cm}
\begin{center} {Fig. 1. The time contour \(C\).}
\end{center} 

\end{document}